\documentstyle[12pt]{article}
\begin{document}

\begin{center}
{\Large

{\bf SHORT RANGE CORRELATIONS IN\\

\vspace{0.3cm}

THE PION S-WAVE SELF-ENERGY\\

\vspace{0.4cm}

OF  PIONIC ATOMS }}
\end{center}

\vspace{24pt}

\begin{center}
L.L. Salcedo$^{1}$, K. Holinde$^2$, E. Oset$^2$ and C. Sch\"utz$^3$
\end{center}

\vspace{18pt}

{\small{
\noindent
1) Departamento de F\'{\i}sica Moderna, Universidad de Granada,
E-18071 Granada, Spain.

\noindent
2) Departamento de F\'{\i}sica Te\'orica and IFIC, Centro Mixto Universidad
de Valencia-CSIC, E-46100 Burjassot (Valencia), Spain.

\noindent
3) Institut f\"{u}r Kernphysik, Forschungszentrum J\"{u}lich GmbH, D-52425
J\"{u}lich, Germany.}}

\begin{abstract}

{\small{
We evaluate the contribution of second order terms to the pion-nucleus
s-wave optical potential of pionic atoms generated by short range nuclear
correlations.
The corrections are sizeable because they involve the isoscalar s-wave
$\pi N$ amplitude for half off-shell situations where the amplitude is
considerably larger than the on-shell one.

In addition,  the s-wave optical potential is reanalyzed
by looking at all the different conventional contributions together:
lowest order, Pauli corrected rescattering term, second order absorptive
effects, terms from the interaction of pions with the virtual pion cloud
(chiral corrections) and correlation effects. Different off-shell
extrapolations for the $\pi N$ amplitude
are used and it is found that, although some individual
terms are sensitive to the extrapolation, the sum of them is rather
insensitive. The results are compared with empirical values from best
fits to the data and are found to be compatible, within theoretical and
empirical uncertainties. The results do not rule out further
contributions but they put stringent constraints on their strength.}}
\end{abstract}

\vfill
\hbox{UG-DFM-34/94\hfill}
\vskip 1cm
\newpage

The original work of Ericson-Ericson on pionic atoms \cite{1} set already
the two largest contributions to the s-wave pion nucleus optical
potential. Its real part is given by
\begin{equation}
{\rm Re}\Pi^{(s)} = - 4 \pi[(1 + \epsilon) (b_0 + \Delta b_0) \rho + (1 +
\epsilon) b_1 (\rho_n - \rho_p) + (1 + \frac{1}{2} \epsilon){\rm Re}B_0
\rho^2]
\end{equation}
with $\epsilon = \frac{m_\pi}{M_N}$, $\rho = \rho_n + \rho_p$ and
\begin{equation}
b_0 = \frac{1}{3} (a_1 + 2 a_3) \,;\quad b_1 = - \frac{1}{3} (a_1 - a_3)
\end{equation}
where $a_1, a_3 $ are the isospin $\frac{1}{2} , \frac{3}{2}$ scattering
lengths given by \cite{2}
\begin{equation}
a_{1} = ( 0.171 \pm 0.004) \, m_\pi^{-1} \,;\quad
a_3 = (-0.105 \pm 0.003) \, m_{\pi}^{-1}
\end{equation}

The magnitude $\Delta b_0$, introduced in \cite{1} and subsequently rederived
in different formalisms with exactly the same results \cite{3,4,5}, accounts
for the Pauli corrected second order rescattering terms. Its derivation
within the many-body framework used in the present paper is shown in
appendix A of ref. \cite{5} and its value is given by
\begin{equation}
\Delta b_0 = - \frac{ 6 k_{F}}{\pi m^{2}_{\pi}} \frac{1}{1 + \epsilon}
(\lambda^2_1 + \lambda^2_2) \,;\quad \lambda_{1,2}
= - \frac{1}{2}m_\pi (1 + \epsilon) b_{0,1}
\end{equation}

In addition, there are other terms of higher order in the nuclear
density, accounted for in terms of ${\rm Re} B_0$, which in ref. \cite{1} were
fit to the data, but which have been subsequently evaluated.
We discuss them below.

One of them is the
real part corresponding to the absorption diagrams, also known as
dispersive correction, which is
calculated in \cite{5,6}. The results were obtained using two
different off-shell extrapolations, one from Hamilton
\cite{7} and the other one from LMM \cite{8}. The uncertainties from
different sources were estimated to be of the order of $30\%$,
the largest one coming from the off-shell extrapolation. Indeed,
the results
obtained there in terms of the second order parameter ${\rm Re}B_0$ were
\begin{equation}
\delta {\rm Re}B_{0,d}
 = 0.032\;m_{\pi}^{-4} \, [5] \,;\quad \delta {\rm Re}B_{0,d} = 0.017
\; m_\pi^{-4} \, [6]
\end{equation}
using the off-shell extrapolations of ref. \cite{7} (Hamilton) and ref.
\cite{8} (LMM) respectively.

The findings in ref.~\cite{9} on the effective density
$\rho_{\rm eff} = 0.5\,\rho_0$ ($\rho_0 = - 0.17$~fm$^{-3}$),
for the s-wave part of the optical potential, allow one to talk
alternatively in terms of an equivalent $\delta b_{0,d}$
parameter, multiplying $\delta B_{0,d}$ by $0.23 \, m_\pi^3$.

On the other hand, recent evaluations of the contribution to the s-wave
self-energy from the interaction of pions with the virtual pion cloud
\cite{10,11}, within the context of Weinberg chiral Lagrangians
\cite{12}, gave rise to a moderate repulsion. These terms are additional
to those considered in \cite{5,6}, which are based on the $\pi N$
interaction alone. Their derivation and interpretation within the
many-body scheme of this paper is given in \cite{11}. In
terms of an equivalent $\delta b_{0,{\rm ch}}$ parameter,
using again the effective density of \cite{9}, one obtains
\begin{equation}
\delta b_{0,{\rm ch}} = - 0.0022  \pm 0.0002 \; m_{\pi}^{-1}
\end{equation}
where the $10\%$ error is an estimate due to second order effects in
the modern chiral perturbation expansion \cite{13}, which improves upon the
Weinberg results.

The present paper adds to these, the effects of short range
nuclear correlations
in conjunction with the  off-shell extrapolation of the $\pi N$ scattering
amplitude.
Short range correlations are taken into  account in the evaluation
of the p-wave pion  self-energy. Their effects are very important
and lead to the famous Lorentz-Lorenz effect \cite{1}. The interpretation
of this effect in terms of nuclear correlations modifying the $\pi +
\rho$ exchange was given in \cite{14}. In spite of its relevance in the
 modification of the p-wave self-energy, a similar work on the modification
of the s-wave self-energy by the nuclear short range correlations has never
been done and this is the purpose of this paper. We follow the ideas of
\cite{14} to perform this work.

In fig.~1 we show diagrammatically how a pion exchange in a rescattering
term is modified due to the simultaneous exchange of other mesons
(in the meson exchange picture) which generate the nuclear correlations.
It is well-known \cite{15} that the effect of this
multiple exchange of mesons, simultaneously with the pion, can be
approximated by means of a correlation function such that, if the diagram of
fig.~1$a$ can be represented by a potential $V(\vec{r}\,)$,
the sum of $1a$ plus
all those of the type $1b$ can be represented by $\tilde{V} (\vec{r}\,)$,
given by
\begin{equation}
\tilde{V}(\vec{r}\,) = V(\vec{r}\,)g(\vec{r}\,)
\end{equation}
where $g(\vec{r}\,)$ is an appropriate correlation function incorporating the
repulsion at short distances. Hence
\begin{equation}
g(\vec{r}\,) = 1 - f(\vec{r}\,) \qquad\hbox{\rm with}\qquad f(0) =1 \,,
\quad f(\infty)= 0
\end{equation}

Since in nuclear matter it is easier to apply Feynman rules in
momentum space, we have instead
\begin{equation}
V (\vec{q}\,) \to \tilde{V} (\vec{q}\,) = \int
\frac{d^3 k}{(2 \pi)^3} V (\vec{q} - \vec{k})\, \Omega (\vec{k}\,)
\end{equation}
where
$$
\begin{array}{ll}
V (\vec{k}\,)  = & \int d^3 r \, e^{i \vec{k}\, \vec{r}} \, V (\vec{r}\,)\\
\Omega (\vec{k}\,)  =  &\int d^3 r e^{i \vec{k}\, \vec{r}}\, g (\vec{r}\,) =
 (2 \pi)^{3} \delta (\vec{k}\,) - B (\vec{k}\,) \, ; \quad
B (\vec{k}\,)  =  \int d^3 r e^{i \vec{k}\, \vec{r}}\, f (\vec{r}\,)
\end{array}
\eqno{(10)}
$$

Hence

$$
\tilde{V} (\vec{q}\,) = V (\vec{q}\,) + V_{c} (\vec{q}\,)\,;\quad
V_{c} (\vec{q}\,) = -\int \frac{d^3 k}{(2 \pi)^3} V (\vec{q} - \vec{k}\,)
B (\vec{k}\,)
\eqno{(11)}
$$

In this way we separate in $\tilde{V} (\vec{q}\,)$ one piece which corresponds
to the ordinary one pion exchange and a second one induced by the correlations.

If we particularize the diagrams of fig.~1 to the s-wave $\pi N$ interaction,
we have the series for the pion propagator depicted in fig.~2. The terms
connected by a one pion exchange line are summed up automatically by the
Dyson equation, by writing the pion propagator which incorporates the proper
pion self-energy. This self-energy corresponds to the irreducible diagrams
(i.e. not connected by a pion line) in figs.~2$b$ and 2$d$.
There we see the lowest order s-wave pion self-energy (2$b$) and a new piece
generated by the correlations (2$d$) where the wavy line corresponds to
$V_c(\vec{q}\,)$ of eq.~(11).

The interesting thing to observe is that, in the correlated potential,
$V$ enters as a function of $\vec{q} - \vec{k}$, with
$\vec{k}$ an integration variable, and it is well-known that the isoscalar
part of the s-wave $\pi N$ amplitude (the one that enters in
fig.~2$d$ for isospin saturated nuclei) is small on-shell due to PCAC,
but grows very fast as one moves to off-shell situations
\cite{7,8,16}.

In order to make a detailed evaluation of this piece let us recall that
the s-wave $\pi N$  $t$-matrix (diagonal in spin)
is written as \cite{17}
$$
- i t_{\beta \alpha} (q, q') = - i 4 \pi \left\{
\frac{2 \lambda_{1} (q, q')}{{m_\pi}} \delta_{\beta \alpha} - i
\frac{\lambda_{2} (q, q')}{m_\pi^2} (q^0 + q'\,^0) \, \epsilon_{\beta
\alpha  \lambda} \tau^{\lambda} \right\}
\eqno{(12)}
$$
with $\alpha$, $\beta$, $\lambda$ pionic isospin indices.
The lowest order pion self-energy corresponding to fig.~2$b$ is written as
$$
\Pi (q) = 4 \pi \frac{2 \lambda_{1} (q, q)}{{m_\pi}} \rho
\eqno{(13)}
$$
and the second order term, induced by
the correlations, corresponding to fig.~2$d$ is given by
$$
\Pi_{c} (q) = - \int \frac{d^3 k}{(2 \pi)^3} \left[ 4 \pi \frac{2 \lambda_{
1} (q, k - q)}{{m_\pi}} \rho \right]^2 \frac{1}{(q-k)^2 - m_{\pi}^2}
 B (\vec{k}\,)
\eqno{(14)}
$$

We observe that the isoscalar amplitude enters there as a function of the
half off-shell variables. For pionic atoms we have
$q \equiv ({m_\pi}, \vec{0}\,)$
and $q - k \equiv ({m_\pi}, - \vec{k}\,)$,  since we are assuming the
exchange of mesons generating the correlations to be static
(a fair approximation if they are of sufficiently short range).

In order to get a quantitative result for $\Pi_{c} (q)$ with some estimation
of the errors we have taken three off-shell extrapolations of the isoscalar
amplitude and three different correlation functions.
The three models used for the extrapolation are those of refs. \cite{7,8}
and the recent one of the J\"ulich group \cite{18}. The first one assumes
for the isoscalar amplitude, of interest here, a
short range piece together with $\sigma$ exchange and has been used in several
many-body calculations \cite{5,19,20}. It gives, at $q^0 = q'\,^0 = m_\pi$,
$$
\lambda_1 (\vec{q} , \vec{q}\,') = - \frac{1}{2} ( 1 + \epsilon)
{m_\pi} \left[ a_{\rm sr} + a_{\sigma} \frac{m_{\sigma}^2}{m_\sigma^2 +
(\vec{q} - \vec{q}\,')^{2}} \right]
\eqno{(15)}
$$
with  $a_{\sigma} = 0.220~m_\pi^{-1}$,
$a_{\rm sr} = -0.233~m_\pi^{-1}$ and  $m_{\sigma} = 550$~MeV.

The second one is based on a separable model for the $\pi N$ interaction
and the use of  dispersion relations. It can be parametrized  in the
region of interest to us as \cite{20}
$$
\lambda_1 (\vec{q}, \vec{q}\,') = \lambda_1(0)(1 + c_1 p^2
+ c_2 p^4 + c_3 p^6 )\exp(-cp^2)
\eqno{(16)}
$$
with $p =|\vec{q} - \vec{q}\,'|$, $\lambda_1(0) = 0.0075$,
$c_1 = 4.98$~fm$^2$, $c_2 = 0.726$~fm$^4$, $c_3 = 0.203$~fm$^6$,
and $c = 0.462$~fm$^2$.

The third model \cite{18} uses (direct and crossed) nucleon and
delta-isobar pole diagrams together with correlated $2\pi$-exchange
in the $\sigma$ ($J$=0) and $\rho$ ($J$=1) channels. It
describes quantitatively all relevant $\pi N$ scattering phase shifts
over the whole elastic region.
We take model (2) of \cite{18} which, for reasons indicated in that
paper, is preferred.  The normalization of the $T$-matrix of \cite{18}
needed for our computation is the one of eq. (1) of \cite{18} with the
equivalence $T^{(s)} \equiv 8 \pi \lambda_1/m_\pi$.
In the region of interest to us this amplitude can be parametrized as
$$
\lambda_1 (\vec{q}, \vec{q}\,') = (l_1 + l_2  p + l_3 p^2)\,
\exp (-(p - p_0)^2/\omega^2)
\eqno{(17)}
$$
with $p = |\vec{q} - \vec{q}\;'|$, $l_1 = 3.254\;10^{-3}$,
$l_2 = 0.1156$~GeV$^{-1}$, $l_3 = 0.1588$~GeV$^{-2}$,
$p_0 = 0.204$~GeV, and $\omega = 0.530$~GeV.
The quantity $\lambda_1$ is plotted in fig.~3 for  the three models.

On the other hand we have chosen three different correlation functions
suited to calculations of nuclear matter:
$$
\begin{array}{lll}
\hbox{\phantom{II}\phantom{III}I)} & g(r) =  1 - j_{0} (q_{c} r) \\
\hbox{\phantom{I}\phantom{III}II)} & g (r) = 1 - \exp(- r^2 / R^2) \\
\hbox{\phantom{I}\phantom{II}III)} &
g (r) = (1- \exp(-r^2/a^2))^2 + b r^2 \exp(-r^2/c^2) \\
\end{array}
\eqno{(18)}
$$
with $q_c = 780$~MeV, $R = 0.75$~fm, $a = 0.5$~fm, $b= 0.25$~fm$^{-2}$ and
$c= 1.28$~fm.

The results for $\Pi_c$
are shown in Table I  where we give the correction in terms of a
coefficient $\delta {\rm Re}B_{0,c}$.
We can see that the contribution from this  correlated piece is
systematically repulsive, and the strength depends  on the
extrapolation and the correlations used.

For a given off-shell extrapolation there is a certain dispersion of the
results using different correlation functions. We use this to obtain an
estimate of the uncertainties. Hence we take the average value and an error
which allows us to  reach the extremes.

The results obtained with the Hamilton and J\"{u}lich
extrapolations are very similar, but they differ appreciably from those of the
LMM model. Although we would favour the results of the J\"{u}lich
model, the large
sensitivity to the off-shell extrapolation is an unpleasant feature. But
we must recall here that the dispersion corrections of eqs.~(5) are
also dependent on the off-shell extrapolation. An interesting finding of the
present work is the realization that there are cancellations between the
two pieces which depend on the off-shell extrapolation, the dispersion
and the correlation corrections, and that when the terms are
added, the sum is small compared to $b_0 + \Delta b_0$ and the dependence
on the off-shell extrapolation is weak. This can be seen in Table II. The
dispersive correction column has been completed with the results of the
J\"{u}lich model using the theoretical framework of \cite{5} \cite{6} and
the J\"{u}lich extrapolation. For this purpose the extrapolation of
$\lambda_2$ is also needed and we find for the region of interest to our
problem
$$
\lambda_2(\vec{q},\vec{q}\,') = \lambda_2(0)(1 + a p^2)
\exp (- b p^2)
\eqno{(19)}
$$
with $p =|\vec{q}-\vec{q}\,'|$, $\lambda_2(0)=0.053$,
$a = 9$~GeV$^{-2}$, $b = 3.9$~GeV$^{-2}$.

In the dispersive results of Table II we have included errors of 20$\%$
due to uncertainties other than the off-shell extrapolation \cite{5,6}.
The sum of the dispersive
and correlation corrections appears in the third column of Table II and
we observe that the results are: i) small, (of the order of 15$\%$ of the
$b_0 + \Delta b_0$ term), ii) rather similar for the different off-shell
extrapolations and iii) compatible within the theoretical errors. We take
an average of this sum and its errors as representative of this
combined contribution and, after transforming it into a $\delta b_{0, d+c}$
term, we show it below, together with the other contributions.

If we add up all the contributions discussed in this paper,
with their corresponding uncertainties, to yield the theoretical
equivalent value, $\bar{b}^{\rm th}_0$ we find:
$$
\begin{array}{lcll}
b_0 & = & - 0.0130 \pm 0.0033 \; m_\pi^{-1}, & \hbox{lowest order}\\[2ex]
\Delta b_0 & = & - 0.0140  \pm 0.0008 \; m_{\pi}^{-1}, &  \hbox{second
order Pauli}\\
& & & \hbox{corrected rescattering}\\[2ex]
\delta b_{0, d+c} & = & + 0.0039 \pm 0.0020  \; m_{\pi}^{-1} ,
& \hbox{dispersion + correlation}\\[2ex]
\delta b_{0, {\rm ch}} & = & - 0.0022 \pm 0.0002 \; m_{\pi}^{-1},
& \hbox{chiral}\\[2ex]
\bar{b}_0^{\rm th} & = & - 0.0253 \pm 0.0063 \; m_{\pi}^{-1}, & \hbox{total}
\end{array}
\eqno{(20)}
$$
The parameter $\Delta b_0$ is evaluated from eq.~(4) using again the effective
density. The errors in $b_0$ and $\Delta b_0$ are those coming from
the experimental errors in $a_1$ and $a_3$.

It is interesting to compare these results to empirical values obtained
from best fits to pionic atom data. Using the potential of Seki and
Masutani \cite{9}, the one of Meirav et al. \cite{21} and the one
of Nieves et al. \cite{22}, and using again the concept of
effective density to write all of them in terms of an equivalent $b_0$
parameter, we find
$$
\begin{array}{ccc}
 b_0 =   & -0.0285  \pm  0.0007  \, m_{\pi}^{-1}  & [9]\\[2ex]
 b_0 =   & -0.0350  \pm  0.0062 \, m_{\pi}^{-1}  & [21] \\[2ex]
 b_0 =   & -0.0323  \pm  0.0007 \, m_{\pi}^{-1}  & [22]
\end{array}
\eqno{(21)}
$$
where the errors are the statistical errors coming from the best fit
(see section 5 of ref. \cite{23} for a reanalysis with these potentials
and the determination of the statistical errors). The results in eqs.~(21)
mean that there are also uncertainties in the empirical value of $b_0$
tied to different assumptions about the shape of the potential.
It is also interesting to observe that although
the theoretical result of eqs.~(20) lies a little
below the empirical results, it is compatible with all of them if
uncertainties are considered. This becomes more obvious if a weighted average
\cite{24} of the empirical results of eqs.~(21) is done, which leads to
$b_0^{\rm emp} = -0.0304 \pm 0.0005 m_\pi^{-1}$, where again the
error is only statistical. The dispersion of results in eqs.~(21)
indicates that
there is an error of the order of $0.002 m_\pi^{-1}$ tied to the
assumptions in the shape of the potentials, such that the empirical value
with its error becomes $b_0^{\rm emp} = -0.0304 \pm 0.0025 m_\pi^{-1}$.
The central value of the theoretical results is 17$\%$ below the empirical
one, but if uncertainties are considered the results are compatible.

One last comment should be made about the role of the $\pi N N$ form
factor in the present calculations. We have used throughout a monopole
form factor with a cut off mass $\Lambda = 1300$~MeV from the full
Bonn $NN$ potential \cite{25}. The form factor affects only the dispersion and
chiral terms. There are indications that the cut off mass, $\Lambda$,
could be smaller \cite{26}. In this case the strength of
both terms is reduced by a similar percentage. Since the
dispersive term is about three times larger than the chiral term, the
net effect is to increase $|\bar{b}^{\rm th}_0|$ and bring it even closer
to the empirical value.

The results of this paper have shown that correlation effects in the
s-wave potential are sizeable and that the
off-shell dependence of the $\pi N$ amplitude must be considered in its
evaluation. The correlation correction is then sensitive to the
off-shell extrapolation used. However, when this correction is added to
the dispersive correction, similarly sensitive to the off-shell behaviour,
the sum of the two effects is small and rather insensitive to the
extrapolation used.

The present results should lead us to reflexion. In the past many
theoretical efforts have been devoted to the search of the ``missing
repulsion'' (see section  4 of ref. \cite{11} for a thorough and
critical discussion of the different theoretical works). The results obtained
here indicate that the ``missing repulsion'', if any, cannot be large.
They  certainly do not exclude other sources of repulsion, like those tied
to the renormalization of pions and nucleons in the inner structure of
the $\pi N$ models \cite{27,28}, but puts severe constraints on their
strength.

\begin{center}
ACKNOWLEDGMENTS
\end{center}
{\sl
This work is partly supported by CICYT contract no.
AEN 93-1205, DGICYT contract PB92-0927, and the Junta de Andaluc\'{\i}a.
One of us (K. H.) wishes to acknowledge the Ministerio de
Educaci\'on y Ciencia for his support in a sabbatical stay at the
University of Valencia.
}

\vspace{2cm}

Table I : Correlation correction to the optical potential parameter
${\rm Re}B_{0} [m_\pi^{-4}]$ (eq. (1))
derived from the pion-nucleon amplitudes of
Hamilton \cite{7} , LMM \cite{8} and the J\"ulich group \cite{18}
using three different correlation functions (eqs.~(18)).

\vspace{0.3cm}

\begin{center}
\begin{tabular}{c|c|c|c|}
$\delta{\rm Re}B_{0, c} [m_\pi^{-4}]$ & I & II & III\\[2ex]
\hline
Hamilton & $-0.0127$ & $-0.0133$ & $- 0.0141$ \\
J\"ulich & $ -0.0051$ & $-0.0146$ & $-0.0094$ \\
LMM & $- 0.0001$  &$ - 0.0080$ & $- 0.0012$
\end{tabular}
\end{center}

\vspace{1cm}

Table II: Dispersive and correlation corrections plus their
sum in terms of the ${\rm Re}B_0$ parameter of eq.~(1).
The second column is the average of the correlation effects of Table I
for the three correlation functions.

\begin{center}
\begin{tabular}{c|l|c|c|}
$[m_\pi^{-4}]$ & $\delta {\rm Re}B_{0, d}$ & $\delta {\rm Re} B_{0,c}$
& sum \\
\hline
Hamilton [7] & $0.032\pm 0.006$ [5] &$-0.013\pm 0.001$ &$0.019\pm 0.007$\\
J\"{u}lich [8] & $0.028\pm 0.006$ &$-0.010\pm 0.005$ & $0.018\pm 0.011$\\
LMM [18] & $0.017\pm 0.004$ [6] &$-0.003\pm 0.005$ & $0.014\pm 0.009$\\
\end{tabular}
\end{center}

\newpage

\noindent
FIGURE CAPTIONS:

\noindent
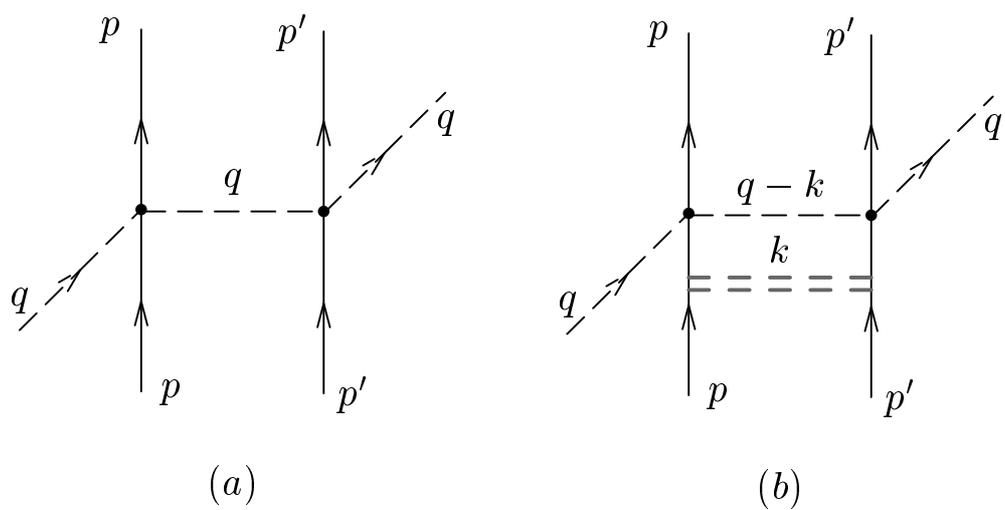
Fig.~1. $(a)$ Second order term in $\pi A$ scattering mediated by pion
exchange alone. $(b)$ Schematic representation of simultaneous
exchange of other mesons together with pion exchange, which are taken
into account in terms of a correlation function.

\noindent
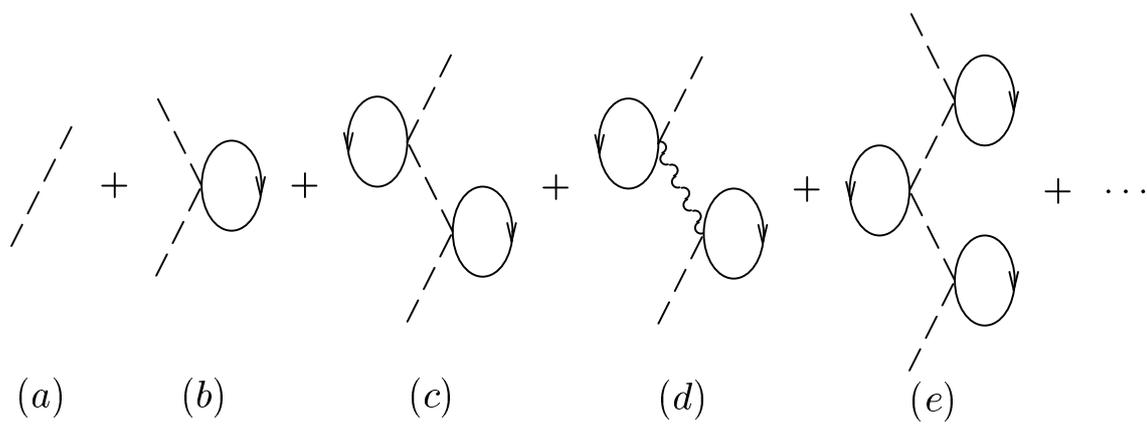
Fig.~2. Diagrammatic representation of the pion propagator led by the
pion s-wave self-energy. The piece $(d)$ is mediated by the correlated
potential of eq.~(11).

\noindent
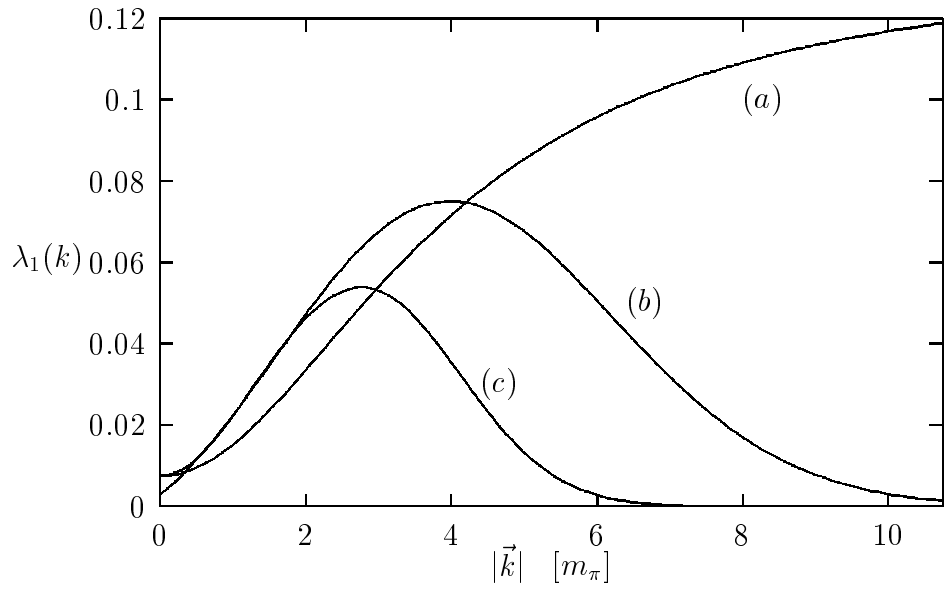
Fig.~3. The value of $\lambda_1(0,\vec{k})$
defined in eqs.~(15), (16) and (17) as a function of the momentum
transfer in units of pion mass, for pionic atom kinematics.
(a) Hamilton model from ref.~\cite{7}. (b) J\"{u}lich group model (2)
from ref.~\cite{18}. (c) LMM model from ref. \cite{8}.

\end{document}